\documentclass[conference]{IEEEtran}
\usepackage{amsmath,amsfonts}
\usepackage{algorithmic}
\usepackage{algorithm}
\usepackage{array}
\usepackage[caption=false,font=normalsize,labelfont=sf,textfont=sf]{subfig}
\usepackage{textcomp}
\usepackage{stfloats}
\usepackage{url}
\usepackage{verbatim}
\usepackage{graphicx}
\usepackage{cite}
\usepackage{hyperref}
\usepackage{multirow}

\begin{document}

\title{Deep Reinforcement Learning-based Video-Haptic Radio Resource Slicing in Tactile Internet}

\author{\IEEEauthorblockN{Georgios Kokkinis, Alexandros Iosifidis, Qi Zhang}

\IEEEauthorblockA{\textit{DIGIT and Department of Electrical and Computer Engineering, Aarhus University} \\
Aarhus, Denmark \\
Email: \{gkokkinis, ai, qz\}@au.dk}
}

\maketitle

\begin{abstract}
Enabling video-haptic radio resource slicing in the Tactile Internet requires a sophisticated strategy to meet the distinct requirements of video and haptic data, ensure their synchronized transmission, and address the stringent latency demands of haptic feedback. This paper introduces a Deep Reinforcement Learning-based radio resource slicing framework that addresses video-haptic teleoperation challenges by dynamically balancing radio resources between the video and haptic modalities. The proposed framework employs a refined reward function that considers latency, packet loss, data rate, and the synchronization requirements of both modalities to optimize resource allocation. By catering to the specific service requirements of video-haptic teleoperation, the proposed framework achieves up to a $25$\% increase in user satisfaction over existing methods, while maintaining effective resource slicing with execution intervals up to $50$ ms.
\end{abstract}


\section{Introduction}
\label{sec::Intro}

The Tactile Internet is envisioned to include the communication of touch, allowing immersive interaction and manipulation of physical or virtual objects across distance. Tactile Internet is expected to create a new dimension for human-to-machine and inter-human interaction, enabling disruptive services and applications such as teleoperation, industrial control, and immersive virtual reality. However, current communication systems fall short of the stringent latency and reliability demands of these applications. Bridging this gap calls for novel solutions, which can integrate edge computing and AI-driven optimization techniques to meet these needs, particularly in transmitting the sense of touch~\cite{Sachs2019}. In Tactile Internet applications, an operator sends a control signal to a remote teleoperator system, which returns force feedback, creating a closed-loop for real-time interaction. Maintaining the stability of this loop requires a high sampling rate of 1 kHz~\cite{antonakoglou2018}. Additionally, the teleoperator must provide both haptic and video feedback, which need precise synchronization between them. Synchronized video-haptic data transmission necessitates a resource management system capable of meeting the distinct service requirements of each modality while ensuring synchronization. 

In Radio Access Network (RAN), the co-existence of haptic and video streams can be achieved through RAN slicing, which logically partitions radio resources to allow independent configuration of each stream~\cite{popovski2018}. In the case of video-haptic teleoperation, both the operator and teleoperator can transmit haptic feedback through an Ultra-Reliable Low-Latency Communications (URLLC) slice, while using an enchanced Mobile Broadband (eMBB) slice for high-bandwidth video feedback. Since the service requirements for video-haptic teleoperation are fixed, they are defined in the RAN-slicing algorithm before deployment and remain constant throughout operation. Resource slicing, however, must be dynamic as the environment of wireless networks is fast-changing. For dynamic RAN-slicing, common heuristics such as splitting resources to equal slices are inefficient as they do not account for slice diversity or complex environment dynamics~\cite{Rongpeng2018}. Determining the optimal resource allocation at any given state is an NP-hard problem, rendering theoretical approaches impractical~\cite{Zhao2019}. As a result, recent research has shifted towards data-driven methods, such as Deep Reinforcement Learning (DRL), where Deep Neural Network (DNN) agents are trained to infer efficient resource scheduling across diverse network conditions. In RAN-slicing, DRL enables the DNN agent to learn through reward feedback based on user satisfaction.

\begin{figure}[!t]
 \label{fig::Overall_teleop}
 \centering
 \includegraphics[width=1\columnwidth]{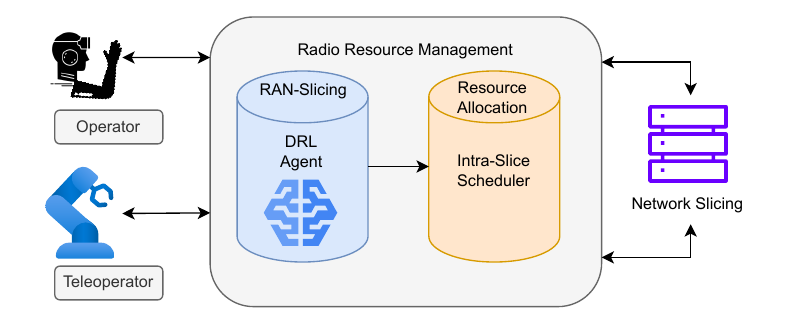}
 \caption{Overall haptic teleoperation pipeline.}
 \vspace{-1em}
\end{figure}

Several DRL-based schedulers have been proposed for radio resource management, primarily targeting general applications and services. In~\cite{Rongpeng2018}, a deep Q-network (DQN) agent collects network observations to allocate resources across video, voice, and URLLC slices, training a DNN to approximate the objective function. Y. Abiko et al.~\cite{Abiko2020} improve this approach by using multiple Dueling DQNs to provide resource scheduling for a variable number of slices. In~\cite{Mei2021}, a high-level DRL scheduler handles inter-slice allocation, while a lower-level scheduler manages intra-slice scheduling at 1 ms intervals~\cite{Chen2023}. However, the use of multiple DRL methods can increase complexity and resource demands.

\begin{figure}
 \label{fig::RAN-slicing}
 \centering
 \includegraphics[width=1\columnwidth]{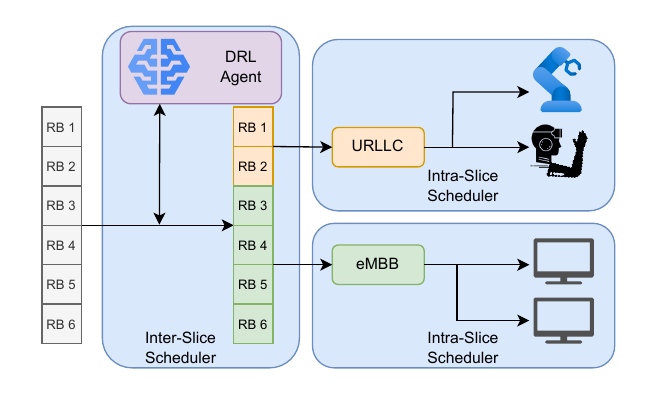}
 \caption{Diagram of radio resource management with RAN-slicing.}
 \vspace{-1em}
\end{figure}

To address the specific Service Level Agreements (SLAs) of network slices, the paper~\cite{Intent2024} prioritizes packet loss and latency for URLLC, and throughput for eMBB using weighted metrics. This approach shows promise for managing the diverse requirements of haptic and video modalities in teleoperation. A traditional reinforcement learning method for haptic data transmission was explored in~\cite{Aijaz2018}, but its Q-learning approach relies on memory-intensive Q-tables. In contrast, DRL uses neural networks, making it more suitable for complex network environments. Moreover, advances in haptic technology, including improved delay tolerance and data reduction through novel encoders, have introduced new service requirements. This paper proposes a framework of DRL-based radio resource slicing that considers video-haptic teleoperation requirements. Simulations demonstrate that the proposed method significantly improves the satisfaction rate for haptic teleoperation services compared to conventional methods. The main contributions of the study are:
\begin{itemize}
    \item Specifying and modeling the video-haptic requirements and synchronization for RAN-slicing frameworks. We use real-world haptic data traces in the performance evaluation.
    \item Implementing a DRL reward function system with enhanced reliability for teleoperation services, which increases the number of satisfied users by up to $25$\% under time-varying channel quality, when compared to conventional DRL methods.
    \item Developing reliable RAN-slicing with inter-slice resource allocation scheduled every 50 ms instead of every 1 ms, which reduces computing costs and is a feasible solution for resource-constrained edge computing.
\end{itemize}

For the remainder of the paper, background technologies related to RAN-slicing and DRL frameworks are outlined in Section \ref{sec::Background}. In Section \ref{sec::DRL-RAN}, the proposed DRL-based video-haptic RAN-slicing framework is detailed, including its mathematical formulations. A comparative analysis with a baseline DRL method is presented in Section \ref{sec::Results}, followed by the conclusions in Section \ref{sec::Conclusion}.


\section{Background}
\label{sec::Background}
This Section provides insight to the key network requirements for haptic teleoperation, as well as an overview of the Deep Reinforcement Learning framework.

\subsection{Teleoperation service slicing}
Bilateral teleoperation consists of at least an operator and a teleoperator. The users (human or machines) are able to constantly exchange haptic data. To achieve immersive teleoperation, the teleoperator also transmits video feedback, either through video streaming or by rendering and transmitting graphics of a virtual environment. In this paper, we consider video transmitted from the remote side. For the haptic and video modalities, two radio resource slices are required. The objective of the RAN-slicing algorithm is to dynamically allocate radio resources to these slices and maximize user satisfaction.
\begin{figure}[!t]
 \label{fig::Haptic_delay}
 \centering
 \includegraphics[width=1\columnwidth]{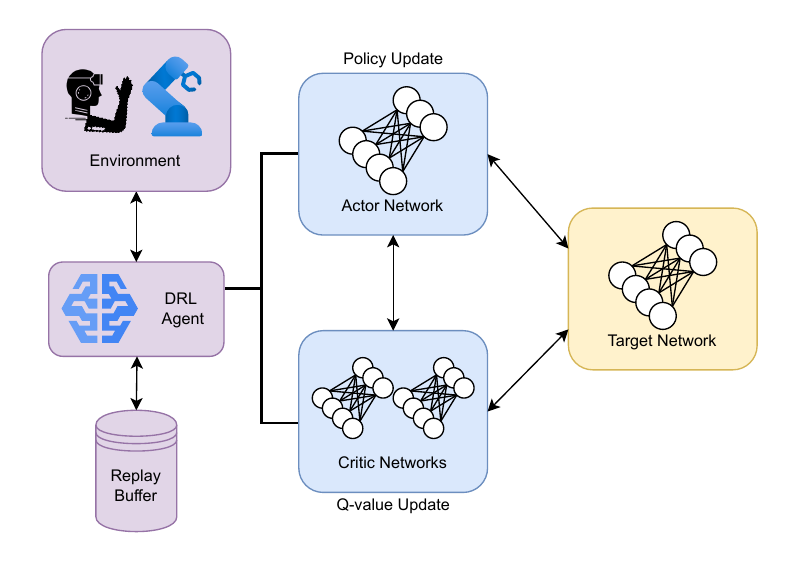}
 \caption{Soft Actor-Critic diagram}
 \vspace{-1em}
\end{figure}

\subsection{Service Requirements}

This study focuses on three primary network requirements for haptic teleoperation: the latency, packet loss, and data rate.

\subsubsection{Latency}
 Low latency is a necessary requirement for haptic signals, as it affects the stability of a closed-loop system. The recommended delay constraint for haptic teleoperation is set according to URLLC at 1 ms~\cite{Tutorial2021}. However, advances of control systems in teleoperation~\cite{XiaoXu2016},~\cite{Sun2016} and human compensation for minor oscillations allow one-way latency tolerance of 10 ms without degradation of task precision or failure~\cite{antonakoglou2018}. If latency exceeds 10 ms, humans will perceive instability and delay in movements, leading to significant degradation in teleoperation performance~\cite{Chaudhari2011}. 
For video feedback, the latency constraint is more relaxed due to lack of stability constraints. However, physiological experimentation indicates that humans detect desynchronization with delays over 50 ms between haptic and video signals~\cite{DiLuca2019}, justifying a stricter video latency threshold. Since we focus on radio resource optimization, we consider buffer latency in the end-to-end (E2E) delay loop.
\subsubsection{Packet Loss}
For packet loss, URLLC standards set a threshold of $10^{-5}$. In modern teleoperation, many redundant haptic packets are omitted due to the use of Perceptual Deadband (PD) coding, which allows the operator to replay the same signal when changes are imperceptible to humans~\cite{Xiao2016}. In contrast, video streaming can tolerate packet loss rates of up to $10^{-1}$.
\subsubsection{Data Rate}
For haptic signals, the baseline requirement for data rate is quite low compared to video streaming, due to the small packet size of haptic signals. With the use of PD coding, we can achieve up to a 90\% reduction in transmitted haptic packets~\cite{Hirche2007}. For video data rate, haptic teleoperation sets a minimum recommendation of 30 Frames Per Second (FPS) with a high definition (HD) camera~\cite{Stotko2019}. For modern video codecs like AV1~\cite{AV1_2018}, 30 FPS can be achieved at a data rate of 4 Mbps.

\begin{table}[!t]
\caption{Notation}
\label{tab:notation_meaning}
\centering
\renewcommand{\arraystretch}{1.15}  
\resizebox{\columnwidth}{!}{
\begin{tabular}{c c c c}
\hline
\textbf{Notation}         & \textbf{Meaning}   & \textbf{Notation}         & \textbf{Meaning}  \\ 
\hline
$h$             & haptic slice &
$\tau$                    & latency    \\
$v$             & video slice &
$r$                       & data rate         \\
$\mathcal{B}$             & system bandwidth   & $\rho$                    & packet loss      \\
$\mathcal{N}_\text{rb}$   & resource blocks &    
$\beta_{\text{o}}$        & buffer occupancy \\ 

$\mathcal{R}$             & DRL reward function &
$U$             & Teleoperation user pairs\\  
$Q$                    & DRL long-term reward & 
$\eta$                    & mean spectrum efficiency
 \\ 
$\mathcal{\pi}$             & DRL action policy   & 
$\sigma$                    & spectrum efficiency fluctuation
\\
$s$             & DRL state  & 
$t$                    & time slot
\\
$a$             & DRL action  & 
$\mathbb{E}$                    & mean value operator
 \\ 
\hline
\end{tabular}}
\end{table}

\subsection{Deep Reinforcement Learning}
In reinforcement learning, an agent selects actions based on the observed state of the environment and refines its policy iteratively through a reward function. For complex scenarios like RAN-slicing with time-varying network conditions, a Deep Neural Network (DNN) approximates the policy function, mapping states to actions. We employ the Soft Actor-Critic (SAC) method, which is particularly effective in dynamic environments. SAC optimizes a balance between the expected return and policy entropy, using an entropy coefficient to manage the trade-off between exploration and exploitation. The agent's experiences, comprising state, action, reward, and next state, are stored in a replay buffer, allowing the SAC agent to sample mini-batches for policy updates~\cite{HaarnojaSAC2018}. Since SAC operates in continuous action spaces, its outputs are mapped to discrete values for resource allocation per slice, as described in~\cite{Intent2024}.


\section{DRL-based Video-Haptic RAN-slicing}
\label{sec::DRL-RAN}
In this Section, we provide an overview of the proposed DRL-based framework for video-haptic RAN-slicing in teleoperation.
\subsection{DRL key components}
The DRL method has the following key components:
\begin{itemize}
\item\textbf{Agent}: The agent is a Deep Neural Network (DNN) trained to optimize the actions taken based on a given observable state of the environment. In SAC, it consists of the actor and critic DNNs, which are responsible for stabilizing and refining the policy updates during training. 
\item\textbf{State:} The network state represents the observable space of the agent and consists of key communication properties of the users, which are the current spectrum efficiency $\eta$, buffer occupancy $\beta_{\text{o}}$, and the latency $\tau$, packet loss $\rho$ and data rate $r$. To reduce the state dimensions, only the average of each component is observed. This can influence the optimization of the DRL's action policy. However, as detailed in subsection \ref{subsec:reward_function}, the latency reward function is determined based on the user with the lowest spectrum efficiency, thereby guiding the resource slicing strategy.
\item\textbf{Action:} In RAN-slicing, the action is the allocation ratio of the total frequency bandwidth per slice. The smallest frequency resource unit in 5G New Radio and 6G is the Resource Block (RB). Therefore, the action of the DRL dictates the number of RBs allocated for each slice, leading to a discrete action space.
\item\textbf{Reward:} The reward feedback is a function that evaluates a given action. In the case of RAN-slicing, the feedback is positive when the actions lead to achieving service requirements for all users. DRL agents are trained to choose actions based on maximizing the long-term reward.
\end{itemize}

\begin{table}[!t]
\caption{Agent Hyperparameters (Soft Actor-Critic)}
\renewcommand{\arraystretch}{1.25}  
\label{tab:SAC_params}
\centering
\resizebox{\columnwidth}{!}{
{\begin{tabular}{c c c c}
\hline
\textbf{Hyperparameter} & \textbf{Value} & \textbf{Hyperparameter} & \textbf{Value} \\ 
\hline
Hidden Layers           & 3              & Batch Size              & 1024           \\ 
Hidden Layer Size       & 512, 512, 256  & Learning Rate           & 0.0001         \\ 
Discount Factor $\gamma$ & 0.99         & Target Update Rate  & 0.005      \\ 
Replay Buffer Size      & $1 \times 10^6$& Entropy Coefficient $\lambda$   & Auto-tuned     \\ 
Optimizer               & Adam           & Target Smoothing         & 0.995          \\ 
\hline
\end{tabular}}}
\vspace{-1em}
\end{table}
\subsection{VH-SAC formulation}
 In RAN-slicing with the soft actor-critic DRL method, the actor-network chooses a slicing action $a_t$ at every time slot t, given a network state $s_t$. The policy $\pi$ with which the actor chooses an action is stochastic, meaning that $a_t$ is selected from a distribution with a probability $\epsilon$. The policy update aims to maximize a combination of the expected rewards and policy entropy:
\begin{equation}
    \pi^{*} = \arg\max_{\pi} \mathbb{E} \left[ \sum_{t} \gamma^t \big( \mathcal{R}(s_t, a_t, s_{t+1}) - \lambda \log \pi_\epsilon(a_t | s_t) \big) \right],
\end{equation}

\noindent where $\gamma$ is the reward discount factor, $\lambda$ is the entropy coefficient. The Q-value, the expected long-term reward, is given by the following equation:
\begin{align}
    &Q_(s_t, a_t) = \mathbb{E} \bigg[\mathcal{R}(s_t, a_t, s_{t+1}) \nonumber \\
    &+ \gamma \big( \min_{i=1,2} Q_{\theta_i}(s_{t+1}, a_{t+1}) - \lambda \log \pi(a_{t+1} | s_{t+1}) \big) \bigg],
\end{align}

\noindent where $Q_{\theta_1} \text{and } Q_{\theta_2}$ are the Q-values of the two critic-networks.

\subsection{Reward function for teleoperation}
\label{subsec:reward_function}
For the reward function $\mathcal{R}$, the key concept is to avoid exceeding the thresholds set by the service requirements for each data modality.
In many studies, the average of each requirement for all users inside each slice is considered. We set $\bar{r}^k,\bar{\rho}^k,\bar{\tau}^k$ to be the average data rate, packet loss, and latency for each slice $k \in \{h,v\}$, where $h$ denotes the haptic slice and $v$ the video slice. For both the haptic and video slices, the data rate reward is calculated as follows:
\begin{equation}
    \mathcal{R}^k_{r} =
    \begin{cases} 
     - \frac{r^k_{{0}} - \bar{r}^k}{r^k_{0}}, & \text{if } \bar{r}^k < r^k_{0} \\
      0, & \text{otherwise}
    \end{cases},
\end{equation}

\noindent where $r^k_0$ is the data rate requirement of the slice. We use a negative reward function in the DRL to encourage minimization of the objective, with the reward reaching zero once the target is reached. Similarly we formulate the reward function for the packet loss:
\begin{equation}
    \mathcal{R}^k_{\rho} =
    \begin{cases} 
      -\frac{\bar{\rho}^k - \rho^k_{0}}{c \rho^k_{0}}, & \text{if } \bar{\rho}^k > \rho^k_{0} \\
      0, & \text{otherwise}
    \end{cases},
\end{equation}

\noindent where we set the constant $c$ at the denominator during the initial stages of training to stabilize the reward contribution of packet loss as it can reach much higher values than the requirement. 

\begin{table}[!t]
\caption{Data traffic and network costraints}
\label{tab:kpi_comparison}
\centering
\renewcommand{\arraystretch}{1.25}  
\resizebox{\columnwidth}{!}{
{\begin{tabular}{c c c}
\hline
\textbf{Traffic characteristic}         & \textbf{Haptic Data} & \textbf{Video Data} \\ \hline
Packet Size (bytes)  & 8                    & 16667              \\ 
Packet Arrival Rate (ms)     & 1                & 33.33            \\ 
Jitter (ms)     & 0                   & [min,max] = [-4,4] \\ \hline

\hline
\textbf{Slice requirement}         & \textbf{Haptic Slice} & \textbf{Video Slice} \\ \hline
Latency (ms)         & $\leq 10$             & $\leq 50$            \\ 
Packet Loss (\%)     & $\leq0.001$                  & $\leq10$                   \\ 
Data Rate (Mbps)     & $\geq 0.2$            & $\geq 4$             \\ \hline
\end{tabular}}}
\end{table}

\begin{table}[!t]
\caption{Simulation Parameters}
\label{tab:sim-params}
\centering
\renewcommand{\arraystretch}{1.25}  
\resizebox{\columnwidth}{!}{
\tiny
{\begin{tabular}{c c c c}
\hline
\textbf{Experiment} & $U$ & $\eta$ (bits / Hz) & $\sigma$ (\%) \\ \hline
Over $U$ & 10 - 25 & 5 & 25 \\
Over $\eta$ & 20 & 4.5 - 5.5 & 25 \\
Over $\sigma$ & 20 & 5 & 0 - 50 \\
\hline
\end{tabular}}}
\end{table}

The latency reward functions for the haptic and video slices $\mathcal{R}^h_\tau, \mathcal{R}^v_\tau$ follow separate calculation approaches.
For the video slice, we consider the synchronization threshold, which is based on the relative delay between haptic and video packets:
\begin{equation}
    \mathcal{R}^v_\tau =
    \begin{cases} 
      -\frac{\left| \bar{\tau}^v - \bar{\tau}^h \right| - \tau_{\text{sync}}}{\tau_{\text{sync}}}, & \text{if } \left| \bar{\tau}^v - \bar{\tau}^h \right| > \tau_{\text{sync}}\\
      0, & \text{otherwise}
    \end{cases},
\end{equation}
where $\tau_\text{sync}$ represents the maximum tolerable synchronization delay between video and haptic packets, ensuring that the relative delay remains within acceptable limits to maintain a seamless user experience.
For the haptic slice, we consider the average values of the data rate and packet loss, but the strict latency requirement must be satisfied for all users. For latency, we calculate the reward according to the user with the worst spectrum efficiency:
\begin{equation}
    \mathcal{R}^h_\tau =
    \begin{cases} 
      - \frac{\hat{\tau} - \tau^k_{0}}{\tau^k_{\text{max}}}, & \text{if } \hat{\tau} > \tau^k_{0} \\
      0, & \text{otherwise}
    \end{cases},
\end{equation}
where $\hat{\tau}$ is the latency of the user with the worst spectrum efficiency, and $\tau^k_{\text{max}}$ is the maximum time that a packet is allowed to stay in the buffer without being discarded. By controlling the haptic latency according to the worst user, we aim for a reliable system for the users who have poor channel quality while maintaining satisfaction of all users.
The total reward function is calculated as the sum of all the components mentioned above:
\begin{equation}
    \mathcal{R} = \sum_{k \in \{h, v\}} \left( \mathcal{R}^k_r + \mathcal{R}^k_\rho + \mathcal{R}^k_\tau \right),
\end{equation}

\begin{figure} [!t]
 \centering
 \includegraphics[width=1\columnwidth]{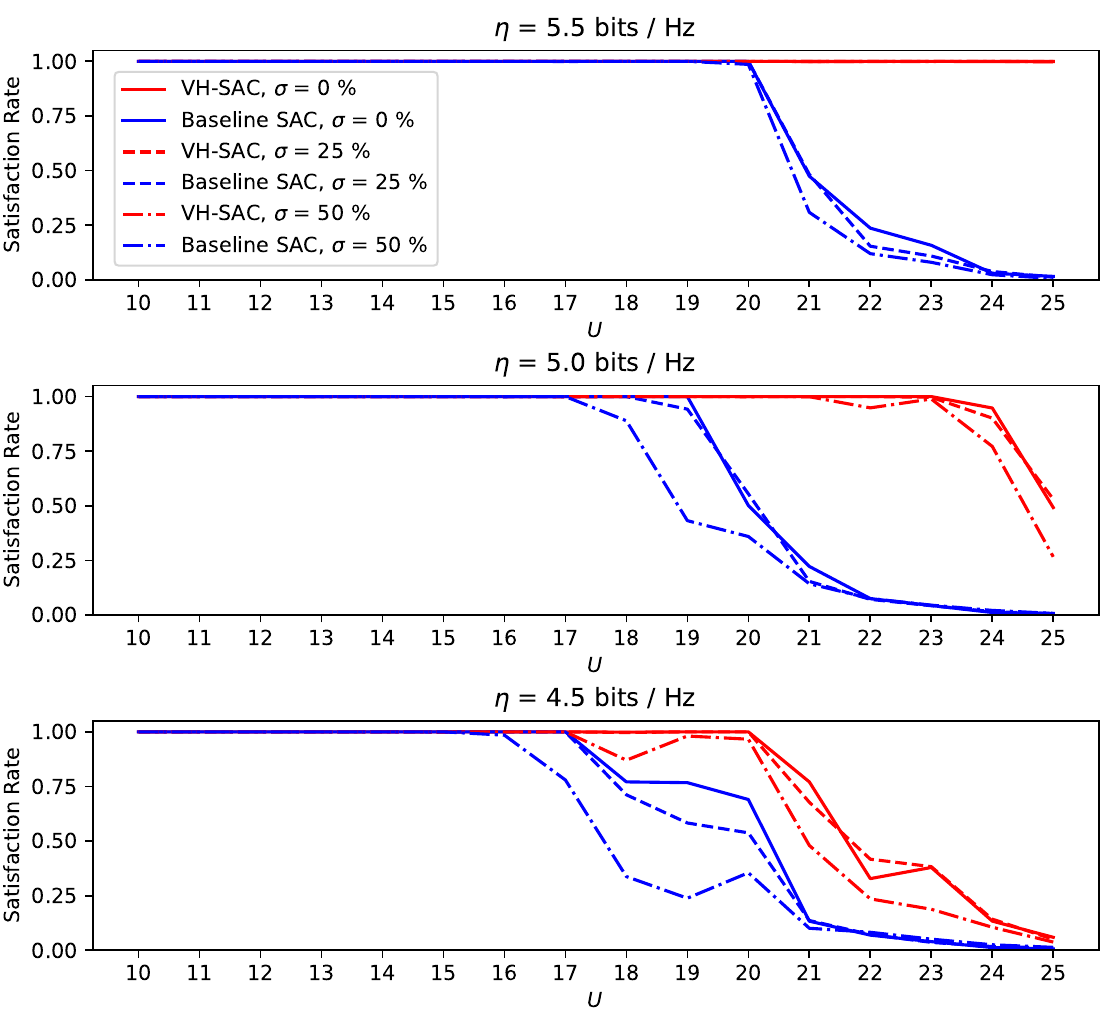}
 \caption{Satisfaction Rate over the number of operator-teleoperator user pairs.}
 \label{fig:SR_Users}
 \vspace{-1em}
\end{figure}

For intra-slice scheduling, we distribute the RBs of each slice $\mathcal{N}^k_{rb}$ to the user RBs $\mathcal{N}^u_{rb}$ proportionally to the user buffer occupancy $\beta^u_{o}$: 
\begin{equation}
\label{eq:scheduler}
    \mathcal{N}^u_{rb} = \left\lfloor \frac{\beta^u_{o}}{\beta_{o}} \mathcal{N}^k_{rb} \right\rfloor ,
\end{equation}
where $\beta_{o}$ is the cumulative buffer occupancy for the whole slice. Unlike conventional resource schedulers, which often use round-robin to achieve uniform resource distribution, the proportional scheduler derived from Eq.~\ref{eq:scheduler} prioritizes users with higher buffer occupancy, allocating more resources to accommodate higher packet loads.


\section{Performance Evaluation}
\label{sec::Results}


\subsection{Baseline}
To the best of our knowledge, no DRL methods have been proposed that are specifically designed to address the latency, packet loss, and data rate requirements associated with video-haptic teleoperation. Hence, we use the SAC method from~\cite{Chaudhari2011} as a baseline, which performs slicing for general URLLC and eMBB services, and adapt it to provide resource allocation for the haptic and video slice of the teleoperation service. This baseline is reasonable for comparison as it utilizes the SAC DRL method, while it also inherits service requirements from URLLC and eMBB standards, yielding better results than common heuristics.

\subsection{Simulation environment}
\subsubsection{Data Traffic}
We consider two types of users for the teleoperation service: the operator and the teleoperator. The operator generates haptic data for position and velocity, while the teleoperator provides haptic force feedback and video feedback. Each teleoperator uses one haptic and one video device, while the operator uses a single haptic device. Haptic data traces were captured from real-world activities using the Novit Falkon haptic device in a virtual environment, including free movements, object tapping, pressing, pushing, and interacting with rigid objects. The link of the database is provided in~\cite{Daniel_traces}. Each haptic packet is 8 bytes (6 bytes for $3$D data and 2 bytes for control overhead). Video feedback uses a statistical model considering an HD camera at 30 FPS. The traffic characteristics and requirements per data modality are summarized in Table \ref{tab:kpi_comparison}.

\begin{figure}[!t]

 \centering
 \includegraphics[width=1\columnwidth]{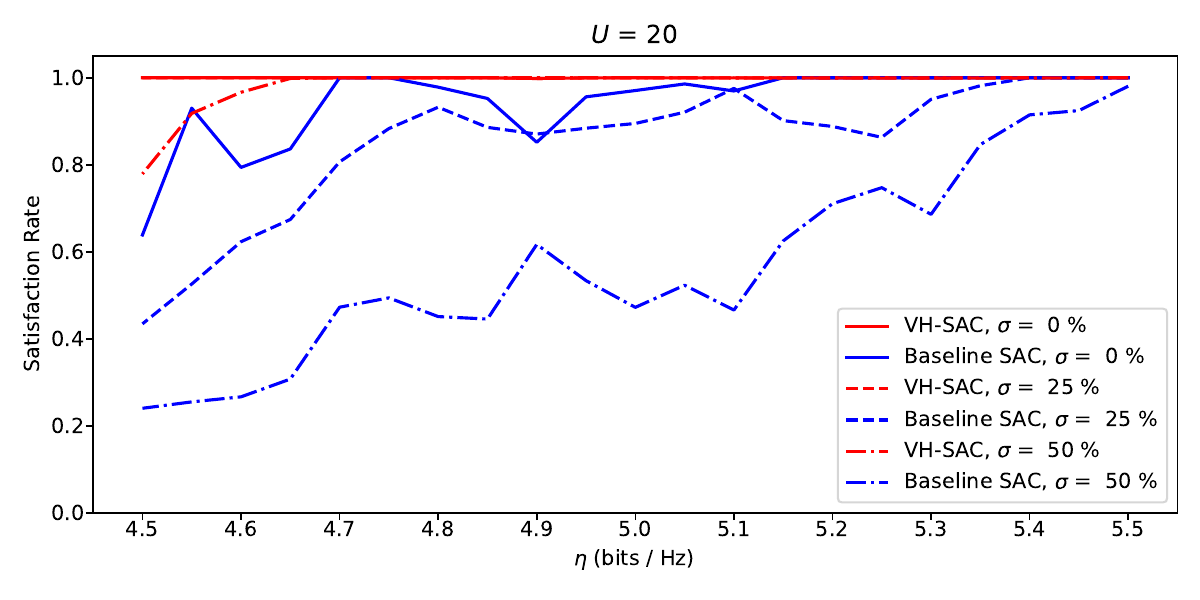}
 \caption{Satisfaction Rate over the mean spectrum efficiency.}
 \label{fig:SR_SE}
 \vspace{-1em}
\end{figure}

\subsubsection{RAN configuration}
We utilize the channel modeling from~\cite{Intent2024}, where Spectrum Efficiency (SE) files are extracted by using Quadriga simulation tool for channel profiling, at Frequency Range 1 with a center frequency of 2 GHz. The files were extracted from users being uniformly distributed during the trials, and the SE is measured every 1 ms. The 1 ms time slot is also used as the Transmit Time Interval (TTI). The provided SE files are time-varying and insert high dynamics to the simulation. In our simulated environment, we consider a bandwidth $\mathcal{B}$ of 20 MHz, divided among $\mathcal{N}_{rb} = 100$ RBs. 
\subsubsection{Simulation Parameters}
For the experiments we iterate over three simulation parameters, the number of operator-teleoperator user pairs $U$ (i.e. for $U = 20$, we have 20 operators and 20 teleoperators), the average spectrum efficiency $\eta$, and the fluctuation of the spectrum efficiency $\sigma$. As shown in Table \ref{tab:sim-params}, we iterate over $U = 10 - 25$, $\eta = 4.5 - 5.5 \text{ bits / Hz}$, and $\sigma = 0 - 50 \text{~\%}$. For each iterated value we keep the other two fixed. 

\subsection{Satisfaction Rate}
\subsubsection{VH-SAC vs. baseline SAC}
\begin{figure}[!t]

 \centering
 \includegraphics[width=1\columnwidth]{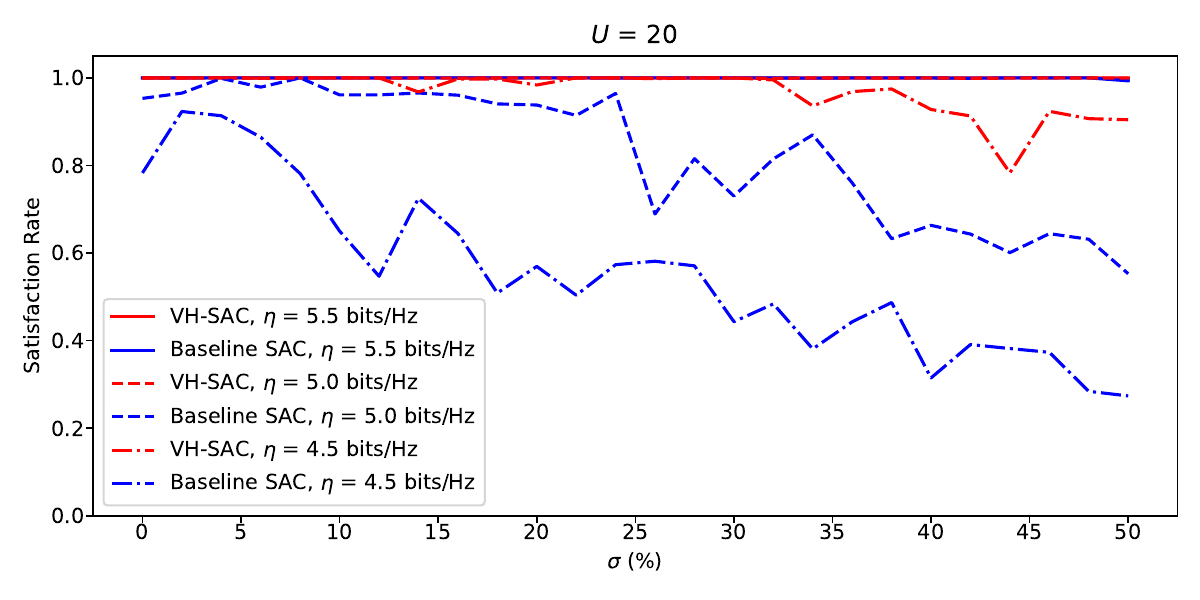}
 \caption{Satisfaction Rate over SE fluctuation.}
 \label{fig:SR_FL}
 \vspace{-1em}
\end{figure}

We evaluate the performance of each method by the satisfaction rate (SR) of the teleoperation users. We define SR as the time that the latency, packet loss and data rate requirements of both operator and teleoperator are satisfied, divided by the total runtime of a simulation trial. Each trial lasts for $1 \times 10^4$ time slots or $10$ seconds. The performance of the proposed video-haptic-aware SAC are shown in Fig.~\ref{fig:SR_Users}, Fig.~\ref{fig:SR_SE}, and Fig.~\ref{fig:SR_FL}. 

In Fig.~\ref{fig:SR_Users}, we measure the SR over an increasing number of pairs of teleoperation users. At average SE of $5.5 \text{ bits} / \text{Hz}$, the proposed method can preserve satisfaction even with fluctuation of $50\%$, while for the baseline SAC the SR drops dramatically when the number of user pairs exceeds 20 (i.e., 20 teleoperation services). While the overall satisfaction rate decreases as the channel quality for all pairs decreases, the proposed method shows much more graceful degradation  compared to the baseline SAC, achieving a higher overall SR. Similar results are observed in Fig.~\ref{fig:SR_SE} and Fig.~\ref{fig:SR_FL}, where we perform simulations over a range of average SE and fluctuation range. Fig.~\ref{fig:SR_SE} shows that for a system of $U = 20$, the proposed method can ensure good SR, while $\eta$ is varying between $4.5$ bits/Hz and $5.5$ bits/Hz, however, the baseline SAC cannot sufficiently support a system of 20 pairs under the same channel conditions, especially during average SE below 5 bits/Hz plus fluctuations. Similarly, Fig.~\ref{fig:SR_FL} shows that in a system of 20 pairs of users, the proposed method can provide stable and high SR in comparison with the baseline SAC under SE fluctuations.

\begin{figure}[!t]

 \centering
 \includegraphics[width=1\columnwidth]{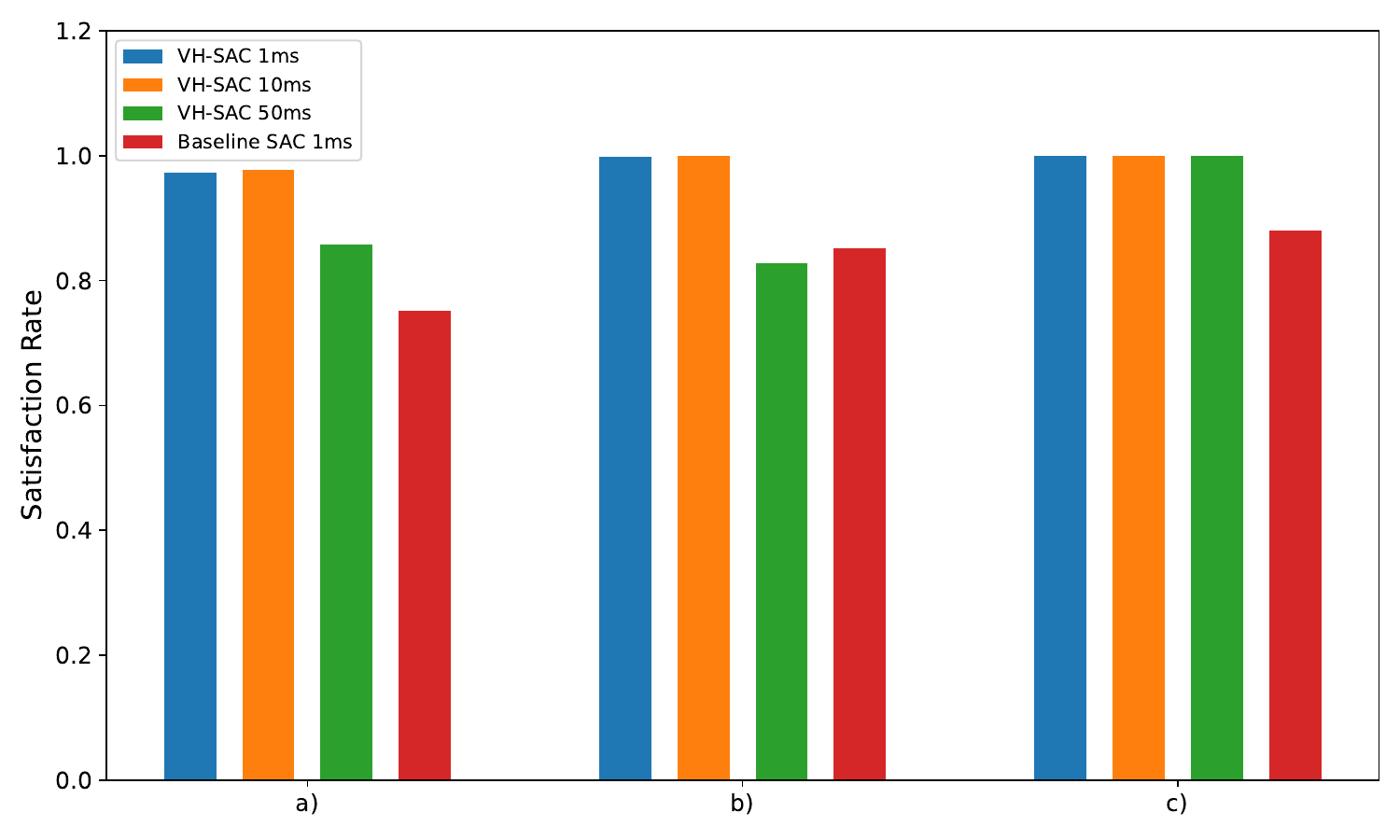}
 \caption{Satisfaction Rate over multi-step prediction and baseline SAC for iterating over a) Teleoperation user pairs $U$ b) mean spectrum efficiency $\eta$ c) SE fluctuation $\sigma$}
 \label{fig:bars}
 \vspace{-1em}
\end{figure}

\subsubsection{Multi-step prediction}
Dynamic RAN-slicing can place substantial demands on computing resources, which becomes especially critical in edge computing scenarios where computational capacity is limited. Frequent updates, such as those occurring every millisecond, exacerbate end-to-end latency as each iteration requires resource-intensive processing. To tackle this challenge, we are interested in measuring the performance of the proposed method when it performs radio resource slicing every 10 or 50 ms, instead of every 1 ms. In Fig.~\ref{fig:bars} we showcase the performance of such scenarios compared to the baseline. While the satisfaction shows minor drop due to decreased scheduling granularity, it is still more robust compared to other methods, even when it runs at every 50 ms. This experiment shows that the proposed method can reduce computation cost in DRL-based resource allocation.

\section{Conclusion}
\label{sec::Conclusion}
In this paper, we developed a DRL-based RAN-slicing framework for video-haptic teleoperation, addressing the unique challenge of managing both video and haptic data streams with a delay synchronization constraint between the two modalities. We utilized real-world haptic data traces for realistic data traffic in the experiments. Our proposed DRL reward function enhances the reliability of video-haptic teleoperation services, significantly improving user satisfaction. Furthermore, our framework remains robust even with extended resource allocation intervals, thus reducing computational overhead and increasing the feasibility of resource slicing deployment at the edge. Our findings therefore contribute to efficient and adaptive resource management solutions, supporting the realization of immersive teleoperation services.
\section*{Acknowledgment}

This research was supported by the TOAST project, funded by the European Union’s Horizon Europe research and innovation program under the Marie Skłodowska-Curie Actions Doctoral Network (Grant Agreement No. 101073465), the Danish Council for Independent Research project eTouch (Grant No. 1127- 00339B) and NordForsk Nordic University Cooperation on Edge Intelligence (Grant No. 168043).

\bibliographystyle{IEEEtran}
\bibliography{bibliography.bib}
\end{document}